\documentclass[letterpaper]{article} 
\usepackage{aaai24}  
\usepackage{footmisc}
\usepackage{times}  
\usepackage{helvet}  
\usepackage{courier}  
\usepackage[hyphens]{url}  
\usepackage{graphicx} 
\usepackage{natbib}  
\usepackage{caption} 
\usepackage{algorithm}
\usepackage{algorithmic}

\usepackage{newfloat}
\usepackage{listings}
\DeclareCaptionStyle{ruled}{labelfont=normalfont,labelsep=colon,strut=off} 
\floatstyle{ruled}
\newfloat{listing}{tb}{lst}{}
\floatname{listing}{Listing}
\title{BertRLFuzzer: A BERT and Reinforcement Learning Based Fuzzer}
\author {
    Piyush Jha\textsuperscript{\rm 1},
    Joseph Scott\textsuperscript{\rm 2},
    Jaya Sriram Ganeshna\textsuperscript{\rm 2},
    Mudit Singh\textsuperscript{\rm 2}, and
    Vijay Ganesh\textsuperscript{\rm 1}
}
\affiliations {
    \textsuperscript{\rm 1}Georgia Institute of Technology, USA\\ \textsuperscript{\rm 2}University of Waterloo, Canada\\
    \{piyush.jha, vganesh45\}@gatech.edu, \{joseph.scott, jsganesh, m286sing\}@uwaterloo.ca
}

\usepackage{algorithmic}
\usepackage{graphicx}
\usepackage{lipsum}
\usepackage{xcolor}
\usepackage{graphicx}
\usepackage{xspace}
\usepackage{multirow}
\usepackage{contour}
\usepackage{float}

\newcommand{\toolname}{\textsc{BertRLFuzzer}\xspace}

\begin{document}

\maketitle

\begin{abstract}
We present a novel tool \toolname, a BERT and Reinforcement Learning (RL) based fuzzer aimed at finding security vulnerabilities for Web applications. \toolname works as follows: given a list of seed inputs, the fuzzer performs grammar-adhering and attack-provoking mutation operations on them to generate candidate attack vectors. The key insight of \toolname is the combined use of two machine learning concepts. The first one is the use of semi-supervised learning with language models (e.g., BERT) that enables \toolname to learn (relevant fragments of) the grammar of a victim application as well as attack patterns, without requiring the user to specify it explicitly. The second one is the use of RL with BERT model as an agent to guide the fuzzer to efficiently learn grammar-adhering and attack-provoking mutation operators. The RL-guided feedback loop enables \toolname to automatically search the space of attack vectors to exploit the weaknesses of the given victim application without the need to create labeled training data. Furthermore, these two features together enable \toolname to be {\it extensible}, i.e., the user can extend \toolname to a variety of victim applications and attack vectors automatically (i.e., without explicitly modifying the fuzzer or providing a grammar). 

In order to establish the efficacy of \toolname we compare it against a total of 13 black box and white box fuzzers: 7 machine learning-based black box fuzzers (DeepSQLi, DeepFuzz, DQN fuzzer, modified versions of DeepXSS, DeepFix, GRU-PPO, Multi-head DQN), 3 grammar-preserving fuzzer (BIOFuzz, SQLMap, baseline mutator), a white box fuzzer Ardilla, a random mutator, and a random fuzzer, over a benchmark of 9 victim websites. We observed a significant improvement in terms of time to first attack (54\% less than the nearest competing tool), time to find all vulnerabilities (40-60\% less than the nearest competing tool), and attack rate (4.4\% more attack vectors generated than the nearest competing tool). Our experiments show that the combination of the BERT model and RL-based learning makes \toolname an effective, adaptive, easy-to-use, automatic, and extensible fuzzer.
\end{abstract}

\section{Introduction}
Over the last few decades, we have witnessed a dramatic increase in the number and complexity of Web applications, and a concomitant rise in security vulnerabilities such as SQL injection (SQLi), Cross-site Scripting (XSS), and Cross-Site Request Forgery (CSRF) to name just a few~\cite{owasp}. Researchers have responded to this problem with increasingly sophisticated and powerful fuzzing tools that range from random mutation fuzzers such as Google's AFL~\cite{afl}, black box grammar-preserving fuzzers~\cite{thome2014search}, to SMT solver-based white box fuzzers such as Ardilla~\cite{kieyzun2009automatic}. 

A common problem developers face is that fuzzers are often required to be grammar-aware, especially when fuzzing applications with sophisticated input grammar. Developing grammar-aware fuzzers is well known to be a time-consuming and error-prone process, since fuzzer developers are required to explicitly provide the input grammar of a victim application and modify/develop the fuzzer appropriately. This is especially cumbersome when developers want to retarget their fuzzers to different applications. An additional issue is that even if one has a grammar-aware fuzzer for a specific application-under-test, the fuzzer may need to be modified as and when newer classes of security vulnerabilities are discovered. For example, a fuzzer aimed at a Web application for discovering SQLi cannot effectively find XSS vulnerabilities. Put simply, fuzzers today do not lend themselves to be easily {\it extended} to newer classes of applications and vulnerabilities in a {\it grammar-aware} manner. 



\subsection{Problem Statement}
More precisely, the problem we address in this paper is to create a Web application fuzzer that takes as input a victim application and a set of seed inputs and outputs a mutation operator that transforms these inputs into an attack vector for a given victim application in an extensible, grammar-adherent, and completely automatic manner. 

That is, we want our fuzzer to have the following properties:
\begin{enumerate}
\item First, the system must be \textbf{extensible}, i.e., the user should be able to extend it to many kinds of attack vectors and victim applications with minimal or no human effort (e.g., by learning from data).

\item Second, it must be \textbf{grammar-adhering}, i.e., the output strings produced by the fuzzer must adhere to the input grammar of the victim application with high accuracy, without requiring that the user supply the input grammar of the victim application.

\item Third, the fuzzer must be \textbf{completely automatic}, i.e., learn or synthesize new mutation operations without being provided labeled data or human modifying the code of the fuzzer (note that, by definition, human-written grammar-adhering mutation fuzzers do not automatically learn new mutation operations). 

\item Fourth, the fuzzer must be \textbf{efficient} (time to first attack must be low) and \textbf{effective} (higher attack rate relative to other state-of-the-art tools). 
\end{enumerate}

\subsection{Machine Learning for Fuzzing}
In recent years, we have witnessed a growing trend of augmenting traditional Web application fuzzing techniques with machine learning (ML) methods, that range the entire gamut of ML training methods from supervised learning \cite{liu2020deepsqli} to Reinforcement Learning (RL) methods \cite{bottinger2018deep}. ML-based fuzzers have a few significant advantages over non-ML-based fuzzers. For example, RL-based fuzzing methods are adaptive, i.e., they can easily be modified to explore different kinds of attack vectors and adapt to a variety of victim applications~\cite{scott2021banditfuzz}. Another advantage is that ML-based fuzzers can learn complex attack vector patterns that may be difficult for humans to detect and hard-code into a non-ML fuzzer. On the other hand, ML-based fuzzers that are based on supervised learning have the disadvantage that they may need a large suite of labeled training data of attack vectors in order to accurately identify patterns that can be converted into mutation operators. Finally, the traditional problem of making mutation fuzzers grammar-adhering also carries over to ML-based mutation fuzzers proposed to date. Depending on the type of victim application and the complexity of their input grammars, specifying such grammars can be expensive, error-prone, and time-consuming. It is preferable that fuzzers learn from data both the grammar of a victim application and the attack vector patterns that are likely to be successful.

\subsection{Brief Overview of \toolname}
To address these challenges, we present, \toolname\footnote{The code can be found at the following URL: https://github.com/bert-rl-fuzzer/fuzzer.git}, a BERT (stands for Bidirectional Encoder Representation from Transformers, part of recently developed powerful Natural Language Processing (NLP) tools such as ChatGPT~\cite{chatgpt}) and RL-based fuzzer. Unlike traditional ML-based and non-ML-based fuzzers, \toolname has all the above-mentioned features. That is, \toolname is automatic, extensible, grammar-adhering, and as our experiments show, it is also efficient and effective.

\subsubsection{Input and Output of \toolname:}
The tool \toolname takes as input a victim application and a list of grammar-adherent seed inputs from a seed generator (these are samples from a known class of vulnerability), and outputs a new attack vector aimed at exposing previously unknown security vulnerabilities in the victim application. 

\subsubsection{Why use a BERT Model?:} As noted earlier, it is well-known that making fuzzers grammar-aware has traditionally been an expensive and labor-intensive process, especially in the context of retargeting fuzzers to different applications and attack vector patterns. Fortunately, the rise of Language Models (LMs) in recent years gives us an amazing opportunity to solve this decades-old problem. The reason is that it is well known that LMs have a surprising capacity to learn grammars of programming languages~\cite{chatgpt,brown2020language,feng2020codebert}, just from some fragments of code without requiring one to specify grammars. Based on this observation, a key insight of our work is that an LM-augmented fuzzer can automatically learn the (relevant fragment of) grammar of victim applications and attack vectors via data (i.e., a set of grammar-adherent attack vectors or seed inputs). This approach of ours has the potential to solve a major problem that fuzzer developers have faced for decades.

\subsubsection{Why use RL in Fuzzing?:} Note that merely learning the grammar of a victim application is not enough, as it may not expose any previously unknown vulnerabilities in a victim application that is being fuzzed. Instead, the fuzzer needs to mutate seed inputs in a way that is highly likely to expose previously unknown vulnerabilities in the victim application. This is a difficult search problem over an exponentially large space of inputs of a victim application. For example, a naive approach would be to modify a seed input, in a grammar-adhering manner, with all possible combinations of an attack vector pattern. However, such an approach suffers from combinatorial explosion.

A better approach is to leverage heuristic search techniques, such as from reinforcement learning (RL) literature, whose goal would be to efficiently zero-in on the victim application's vulnerabilities. Put differently, the advantage of a properly designed RL technique is that it can often explore and learn an attack vector pattern that is specific to a given victim application, and do so efficiently and in a completely automatic manner. Based on this general principle, \toolname has a stateful RL agent that learns a mutation operator (i.e., a sequence of operations that takes as input a suitable grammar-adherent input and converts it into an attack vector for a given victim application). Since this learning occurs via feedback from the given victim application (that may contain regular expression sanitizers to protect the application), \toolname is able to find weaknesses in the said application, if they exist. The mutation operators learned via RL in \toolname enable it to automatically explore a space of attack vectors in a heuristic way specialized to a victim application, without any human intervention, rendering \toolname automatic.

\subsubsection{Putting BERT and RL Together in \toolname:} In \toolname, a pre-trained BERT model (trained on seed inputs of a victim application, thus enabling it to learn the grammar of said application as well as attack patterns) acts as the agent in an RL loop that is interacting with the victim application. The RL loop in turn enables \toolname to zero-in on those mutations that are highly likely to expose security vulnerabilities in the victim application.

The ability of BERT models to learn a representation of grammars also enables \toolname to be easily extensible, i.e., with an appropriate data set of unlabelled seed inputs, the user can retrain the BERT model to generate a new class of mutation operators for a given victim application. The RL loop then fine-tunes this pre-trained BERT model by leveraging the victim application, the mutation operators, and the reward mechanism to search through a space of inputs in order to generate a new class of attack vectors for the given application. What makes our approach particularly appealing is that the user does not need to encode grammar for attack patterns. All of this is automated for them via the use of a BERT model and an appropriately designed RL loop. 

Using RL to mimic the behavior of an actual adaptive attacker is now widely accepted \cite{bottinger2018deep, erdodi2021simulating, zhou2021autonomous}. However, to the best of our knowledge, using the BERT model as an RL agent in a fuzzer is novel. It ensures that our fuzzer's RL agent makes a good judgment (hence, pruning down the search space) based on the learned syntax and semantics (just like a hacker) instead of suggesting random mutations that may not guarantee that the output strings are grammar-adherent.

To properly evaluate the scientific merit of our ideas, we perform an extensive and thorough experimental comparison of \toolname with 13 other black box, white box, ML, and non-ML based fuzzers over a curated benchmark of 9 victim Web applications that range in size from a few hundred to 16K lines of code. Through our research questions (Section~\ref{eval}), we show that \toolname has all the properties we require of an effective, efficient, grammar-adhering, extensible, and automatic modern ML-based fuzzer.

\subsection{Contributions}
\begin{itemize}
  \item We present \toolname, a novel BERT and reinforcement learning-based Web application fuzzer that is \textbf{automatic} (learns mutation operators without human assistance), \textbf{extensible} (can be extended to newer classes of security vulnerabilities and victim applications), \textbf{grammar-adhering} (outputs attack vectors that adhere to the grammar of victim applications with high accuracy), \textbf{effective} (is able to find more security vulnerabilities than competing state-of-the-art tools), and \textbf{efficient} (time to first attack is low). To the best of our knowledge, no other ML-based fuzzer uses a BERT architecture and an RL-based algorithm to solve the above-stated problem.
  
  \item We perform an extensive empirical evaluation of \toolname against a total of 13 black box and white box fuzzers: 7 machine learning-based black box fuzzers (DeepSQLi, DeepFuzz, DQN fuzzer, modified versions of DeepXSS, DeepFix, GRU-PPO, Multi-head DQN), 3 grammar-adhering fuzzers (BIOFuzz, SQLMap, baseline mutator), a white box fuzzer Ardilla, a baseline random mutator, and a baseline random fuzzer. We validated the efficacy and efficiency of \toolname over a benchmark of 9 victim websites with up to 16K lines of code. We observed a significant improvement in terms of time to first attack (54\% less than the nearest competing tool), time to find all vulnerabilities (40-60\% less than the nearest competing tool), and rate of vulnerabilities found (4.4\% more than the nearest competing tool) over a variety of real-world benchmark websites.
\end{itemize}

\section{Background}

\noindent{\bf Transformers and BERT Models:} \label{bert_mlm_backgr} Transformer-based models \cite{vaswani2017attention} are critical components in the hugely successful Natural Language Processing models such as GPT-3~\cite{brown2020language}, and ChatGPT \cite{chatgpt} and Google's PALM~\cite{chowdhery2022palm}. More recently, they have been successfully applied in the context of formal programming language applications such as code translation \cite{lachaux2020unsupervised, mastropaolo2021studying}, code synthesis \cite{allamanis2013mining,chen2021evaluating}, and code understanding \cite{mou2016convolutional,guo2020graphcodebert,feng2020codebert}. 

Bidirectional Encoder Representations from Transformers (BERT) models are based on the Transformer architecture~\cite{devlin2018bert, liu2019roberta}. BERT models take as input (textual) strings over some finite alphabet and encode them into a vectorized representation.  For more details, we refer the reader to the paper by Bommasani et al. for a comprehensive overview of BERT models~\cite{bommasani2021opportunities}, and additionally we refer to Devlin et al.~\cite{devlin2018bert} and Liu et al.~\cite{liu2019roberta}.

\vspace{0.2cm}
\noindent{\bf Reinforcement Learning:} There is a large literature on RL and Proximal Policy Optimization (PPO), and we refer the reader to the excellent book by Sutton and Barto~\cite{sutton2018reinforcement} for further reading. 
Deep Q Network (DQN)~\cite{mnih2015human} is a stateful RL algorithm that utilizes deep neural networks to approximate the Q-value function (i.e., the optimal expected long-term reward), allowing for the estimation of the optimal action to take in a given state. PPO~\cite{schulman2017proximal} combines value-based and policy-based methods to optimize policies by using a trust region optimization approach to update them toward better actions. Multi-Arm Bandit (MAB)~\cite{sutton2018reinforcement} is a stateless RL algorithm that involves balancing the exploration of different options (arms) with the exploitation of known, high-reward options in order to maximize the cumulative reward over time.

\vspace{0.2cm}
\noindent{\bf Software Fuzzing:} Software fuzzing is a vast, impactful, and active research field in software engineering, and we refer the reader to Manes et al.~\cite{manes2018art} for a recent survey of the field. The fuzzing terms we use in this paper are standard. 

\vspace{0.2cm}
\noindent{\bf Grammar-adhering Fuzzing, Attack Patterns, Victim Application:} We introduce the following new terms in this paper. The term {\it grammar-adhering fuzzer} refers to a computer program that takes as input a string and outputs a string that adheres to the grammar of a victim application, with high accuracy. Observe that this is a non-standard definition, that encompasses traditional error-free human-written grammar-preserving fuzzers as well ML-based fuzzers that may learn with high accuracy a suitable representation of the grammar of a victim application. The term {\it grammar-adherent mutation operator} refers to a program that outputs a modified input string such that it adheres to the grammar of a victim application with high accuracy. The term \textit{attack pattern} refers to sub-strings of an attack vector (e.g., tautology patterns in SQLi). We use the term {\it victim application} synonymously with application-under-test.

\section{A Compelling Use case of \toolname}

A particularly strong use case of our tool \toolname is for developers of victim applications (e.g., Web application) that have complex input grammars and use sanitizers that have unknown vulnerabilities, and where developing or updating human-written grammar-preserving mutation fuzzers may be expensive. Further, the developers may have sample inputs for a particular class of vulnerabilities, but such a test suite may not be comprehensive, and thus, developers might miss interesting variants of certain attack vectors for which their sanitizers and/or application have no defenses.

In recent years many grammar-preserving mutation fuzzers have been developed for a variety of victim applications and classes of security vulnerabilities~\cite{thome2014search, erdodi2021simulating}. However, such tools need to be reprogrammed by a human every time a new class of security vulnerability is discovered or if they are repurposed for a previously unseen class of victim applications. Further, victim applications might be protected by error-prone sanitizers that could give Web application developers a false sense of security. Finding weaknesses in such sanitizers is particularly important if our goal is to improve the security of the Web ecosystem in general. Hand-written fuzzers have to be modified with knowledge of the weaknesses of a given sanitizer. As these sanitizers are changed by developers, they can introduce newer vulnerabilities. Once again, hand-written fuzzers have to be changed in response. All of this can be very time-consuming and expensive. 

One way to solve the above-described problem is via a grammar-adhering mutation fuzzer, that is automatic and extensible (aka, adaptive) to novel classes of vulnerabilities, victim applications and sanitizers. It is also very important that the fuzzer be automatic, i.e., learn a useful representation of the grammar of victim applications and attack patterns without requiring the human to specify grammars or pattern recognizers (e.g., regular expressions). 


Our tool \toolname provides all the appropriate features for the above-described use case (See Figure~\ref{fig:arch_diag} for architectural details). \toolname learns a useful grammar representation for a given victim application/sanitizer combination (thanks to our use of BERT models), and thus produces grammar-adherent mutation operators, which in turn guarantees grammar-adhering attack vectors. The RL loop enables our tool to probe the victim application's weaknesses, and thus heuristically and efficiently search through a combinatorial explosion of possible variants of a class of attack vectors that are likely to succeed. The use of the BERT model and the RL loop together makes the entire process automatic. Our tool not only looks at the possible variations of a specific class of attack vectors but also can easily be adapted to other classes of security vulnerabilities (provided we have a good set of seed inputs to learn), making it extensible.

\subsection{Example of an Attack}

\begin{figure}[t]
  \centering
  \includegraphics[width=\linewidth]{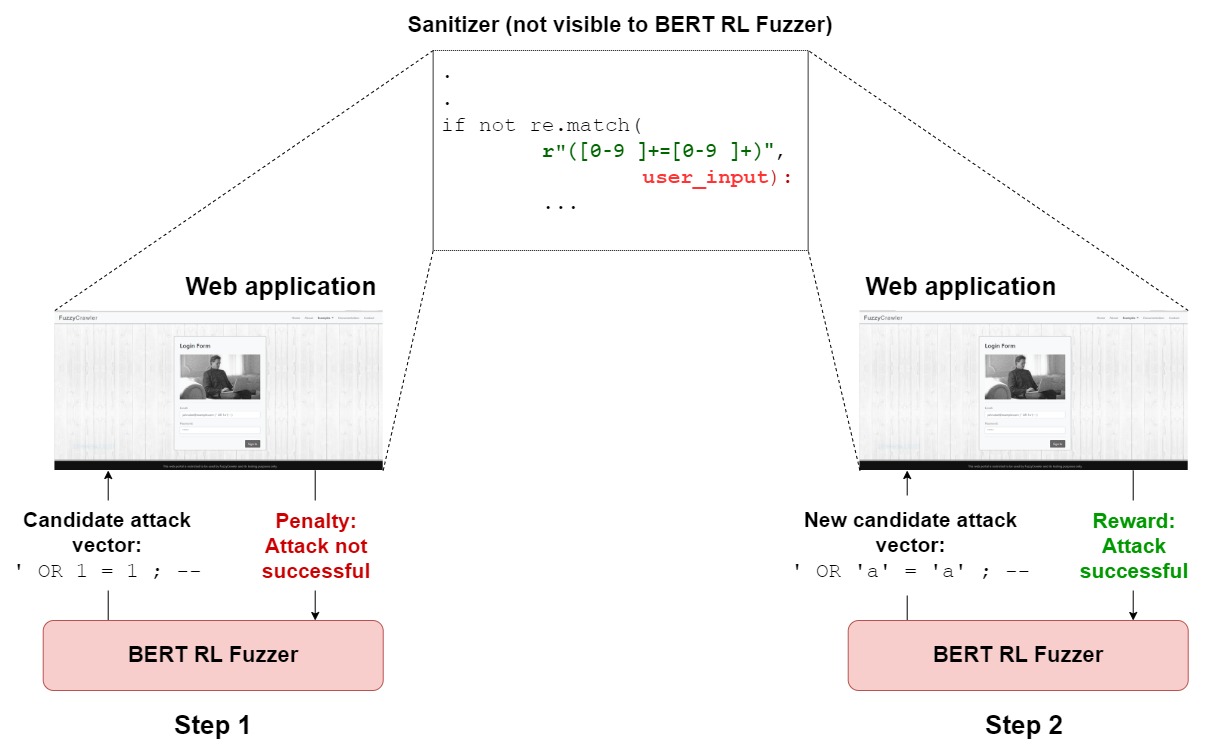}
  \caption{Example of an attack}
  \label{fig:example}
\end{figure}

Let us consider the illustrative, but simple, example in Figure \ref{fig:example}. We test a victim website for the presence of SQLi vulnerabilities. The website has a human-defined sanitizer that rejects the user input for a tautology pattern where two numbers are checked for equality. Unaware of this sanitization in the Web application, our tool generates a candidate attack vector \texttt{' OR 1 = 1 ; -{}-} after modifying a certain seed input, and sends it to the testing environment. As expected, the sanitizer rejects the input, and the RL agent receives a penalty for choosing that mutation operation. In the next step, the RL agent tries a different mutation operation. It replaces \texttt{1 = 1} with \texttt{'a' = 'a'} to generate a new grammar-adherent candidate attack vector: \texttt{' OR 'a' = 'a' ; -{}-}. This time, the input string is able to bypass the sanitizer and result in a successful attack. Note that the \toolname was never trained on the seed input - \texttt{' OR 'a' = 'a' ; -{}-}. \toolname learned the token \texttt{'a' = 'a'} using a different seed input (\texttt{IF ('a' = 'a') THEN dbms\_lock.sleep(5); ELSE dbms\_lock.sleep(0); END IF; END;}) and was able to adapt it for a tautology attack. 

In other words, the \toolname works well even if the seed inputs only reflect part of the input grammar. The seed input in this example does not have the tautology attack pattern with strings (e.g.,' OR 'a' = 'a'; -{}-), instead it only has a tautology attack pattern with numbers (e.g.,' OR 1 = 1 ; -{}-). However, as shown in Figure 1, the model is still able to come up with the string-based tautology attack. The RL loop allows the model to explore different mutation operations, and it chooses 'a' = 'a' because it is part of the vocabulary of the BERT model. The advantage of automatically searching through a space of attack patterns, over a human modifying attack vectors, is that such a process makes it easier to identify potential vulnerabilities that a typical software developer might miss.

\section{\toolname}
\begin{figure*}[ht]
  \centering
  \includegraphics[width=\linewidth]{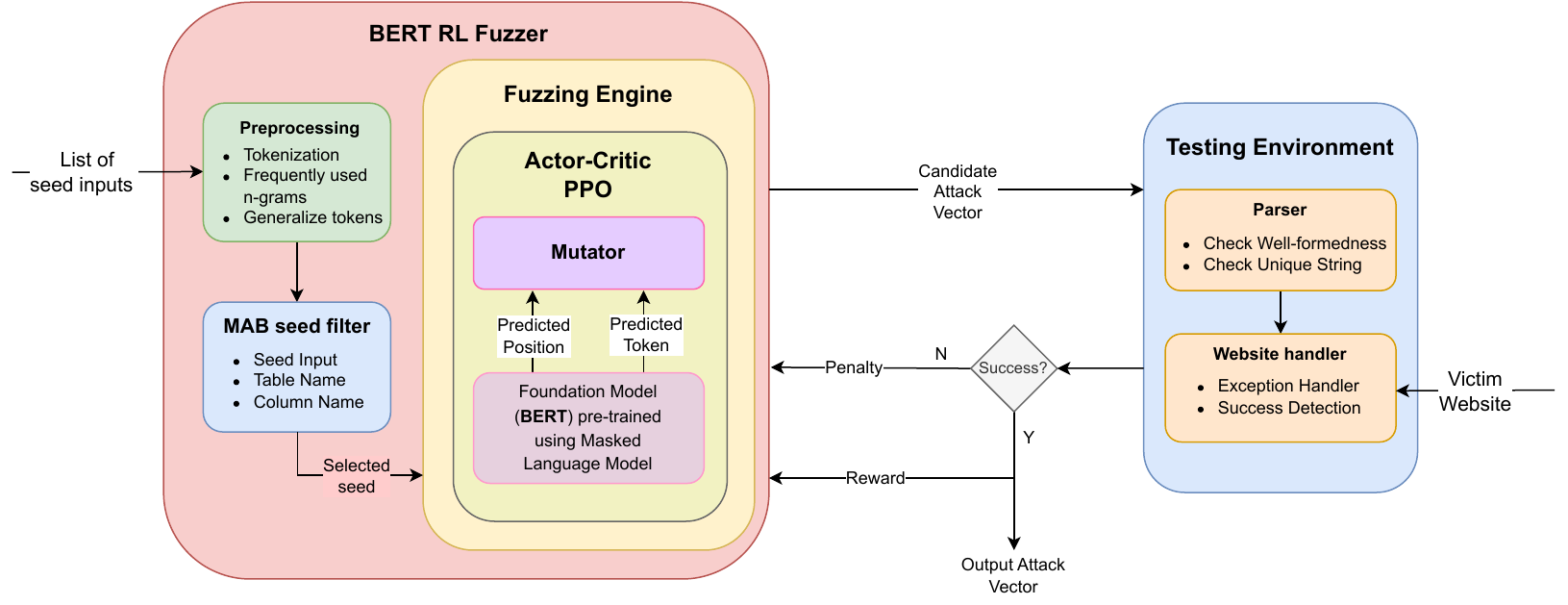}
  \caption{Architecture Diagram of \toolname} 
  \label{fig:arch_diag}
\end{figure*}

In this Section, we present an overview of the input-output behavior and details of the inner workings of our tool \toolname (See also Figure~\ref{fig:arch_diag} for architectural details).

\subsection{Input and Output of \toolname}
Given a victim application $A$ and a set $S$ of seed inputs, \toolname outputs a grammar-adhering mutation operator and a concomitant attack vector (i.e., the mutation operator mutates a seed input into an attack vector) for the given victim application $A$. It is possible that the seed input may be mutated multiple times by different mutation operator output by \toolname before it is deemed to be an attack vector.

\noindent {\bf List of seed inputs:} The user provides a list of grammar-adhering seed inputs that are samples from a known class of attack vectors (easily found in the public domain) for a given victim application. The seed inputs must be representative of the vocabulary and grammar of some known attack vectors (to better illustrate this point, note that \toolname cannot come up with a UNION-based attack if it has not seen one before). These common attack patterns would be used as the pre-training dataset for the BERT model and seed inputs for the \toolname. The goal here is for the tool \toolname to learn a meaningful representation of the input grammar and attack patterns and then automatically generate appropriate attack vectors specialized for a given victim application. We gathered over 7K attack vectors to pre-train the BERT model from the files in public GitHub repositories focused on injection attacks. We follow the pre-training steps as mentioned in Liu et al.~\cite{liu2019roberta} and Huggingface documentation~\cite{wolf2019huggingface}.

\noindent {\bf Victim Application:} A Web application under test that is being checked for the presence of security vulnerabilities. Note that the RL technique implemented in \toolname enables it to generate attack vectors that are specific to a particular victim application/sanitizer.

\noindent {\bf Attack vector:} For a given victim application, the term attack vector refers to any input string that can potentially exploit security vulnerabilities in the said application. 

\subsection{Architectural Details of \toolname}

\noindent {\bf Preprocessing:} In this step, the list of seed inputs is appropriately tokenized using a standard NLTK\cite{nltk} tokenizer. Frequently used n-grams are also added to the vocabulary list. Lastly, SQL table and column names are replaced with generalized tokens to prevent noisy pre-training~\cite{liu2020deepsqli}.

\noindent {\bf Multi-armed Bandit (MAB) Seed Filter:} In this step, we use an MAB agent with Thomson sampling to select a seed input from the list of preprocessed seed inputs. This RL agent, different from the main RL loop of \toolname, helps prevent the tool from getting stuck in local minima, which is a major problem in fuzzing in general ~\cite{saavedra2019review, duchene2013fuzz, gerlich2020optimizing, manes2020ankou}. As training progresses, the MAB agent learns to choose the seed input that results in a higher chance of generating a successful attack vector. (Note that we do not claim the MAB seed filter as one of our contributions.)

\noindent {\bf Actor-Critic Proximal Policy Optimization:} The selected seed input is now passed on to the Actor-Critic Proximal Policy Optimization (PPO) agent of \toolname. We use a pre-trained BERT model (pre-trained on a set of seed inputs using Masked Language Model objective \cite{devlin2018bert, liu2019roberta}, a standard approach used for BERT models) to serve as the building block of the Actor-critic agent. The BERT model is fine-tuned by the RL component to classify the appropriate mutation operations. The BERT model encodes the tokenized input string into a vectorized representation. One can view the actions of the agent as a pair of two sub-actions, where the first is selecting the position in the tokenized list that must be mutated (inserted/deleted/replaced), and the second is choosing the appropriate token that needs to be replaced at that position. The RL agent selects the actions by sampling from a probability distribution. These actions are provided to a mutator to generate a candidate attack vector sent to the testing environment. The RL algorithm is stateful, and the state represents the seed input passed to the BERT agent of the PPO RL model. We set the gamma parameter of the PPO agent to 0.99, 3e-5 as the policy learning rate, and 1e-3 as the value function learning rate and use an Adam optimizer. 

\subsection{Testing Environment}
We create several Web environments with different Web pages vulnerable to various SQLi and XSS attacks. Most Web pages also include input validation checks and sanitizers (e.g., using regex). These Web pages serve as the environment for training our \toolname algorithm. 

We use our custom crawler to parse the Web pages and extract the injection points (e.g., user input fields). An adapter library has been created to send out the candidate attack string to the testing environment. This library is also responsible for sending feedback to the agent corresponding to whether the last test input resulted in a successful/failed attack and whether it could parse successfully. We use the standard methods used by the previous researchers (in DeepSQLi~\cite{liu2020deepsqli} and Ardilla~\cite{kieyzun2009automatic}) to detect the vulnerabilities and verify their presence by manually inspecting the code.

\subsection{Details of Reward and Penalty} \label{rew_pen}
The BERT and the MAB RL agents in \toolname receive a discrete reward signal from the environment in case of a successful SQLi or XSS attack. However, this reward is sparse and the algorithm does take a lot of time to learn if we only rely on this binary feedback. Instead, we introduce different (discrete) penalties for unsuccessful cases. Once the fuzzing engine has generated the new mutated string, we feed it into a parser to check whether the resulting string is a well-formed statement or not. If not, we penalize the agent for the last actions. Furthermore, to generate a successful string in the least number of mutation steps, we give a small penalty if the string passes the parser check, but the adapter said it was not a successful attack. Lastly, we introduced a penalty for generating a previously observed string to encourage more unique errors. (Note that we could have leveraged the parsers that victim applications must have, as indeed an application developer and user of \toolname is most likely to do. The only reason we wrote a parser is because we didn't want to modify the victim application and introduce new errors or break it in any way.)

\subsection{Putting it all together: How \toolname Works} 
Initially, \toolname's seed generator generates samples from a list of seed inputs (these are samples from a known class of vulnerability) that are preprocessed and tokenized. The MAB agent filters and selects one seed input from the list and feeds it to the pre-trained BERT Actor-Critic PPO agent. The PPO agent predicts a grammar-adhering mutation operator, that is then used to create a candidate attack vector. This candidate vector is passed on to the victim website in the testing environment and sends a reward/penalty signal back to the RL agents. Success is detected when the purported attack vector is indeed able to launch an attack on the victim application. In case of a reward (or penalty), the RL agents modify the probability distribution to prefer (or reject) that mutation operation for that state. The previously mutated candidate vector is now used as an input for the PPO agent to predict new mutation operations. The loop is repeated several times before discarding the current input string and choosing a new one from the MAB agent. The fuzzing process terminates when the desired timeout or epoch has been reached.

\section{Experiment Setup} \label{expt_setup}
\subsection{Competing Fuzzers} 
 We compare \toolname against a total of 13 black box and white box fuzzers, including different ML and non-ML based fuzzers. The various machine learning-based black box mutation fuzzers used for our experiments are:

\begin{itemize}
\item \textbf{DeepSQLi} \cite{liu2020deepsqli}: Translates user inputs (or a test case) into a new test case, which is semantically related and potentially more sophisticated due to the ability to learn the "semantic knowledge" embedded in SQLi attacks. A sequence-to-sequence Transformer network is trained on a manually created grammar-abiding training dataset to replicate the mutation operations and generate better attack-provoking vectors. 
\item \textbf{DeepFuzz (modified)} \cite{liu2019deepfuzz}: A Recurrent Neural Network (RNN), that was modified by us to support SQLi and XSS attacks using Deep Q-Network (DQN). The model predicts the mutation operator (i.e., a sequence of operations that takes as input a suitable grammar-adhering benign input and converts it into an attack vector for a given victim application).
\item \textbf{DQN fuzzer} \cite{zhou2021autonomous, erdodi2021simulating}: A Deep Q-Network model that predicts mutation operations specific for certain attack patterns. 
\item \textbf{DeepXSS (modified)} \cite{fang2018deepxss}: Designed for XSS classification using Long short-term memory (LSTM) networks. We modified it to support fuzzing using DQN to predict the mutation operator re-using the existing LSTM networks.
\item \textbf{DeepFix (modified)} \cite{gupta2017deepfix}: Modified the DeepFix tool to use its Gated Recurrent Unit (GRU) component as a fuzzer for both SQLi and XSS using DQN to predict the mutation operator. 
\item \textbf{GRU-PPO Fuzzer}: Created a Gated Recurrent Unit (GRU) based Proximal Policy Optimization (PPO) RL agent to predict the mutation operator. 
\item \textbf{Multi-head DQN Fuzzer}: Created a Multi-head self-attention-based DQN agent to predict the mutation operator. 
\end{itemize}

\noindent The various human-written grammar-adhering fuzzers (where humans explicitly specified the grammar of victim applications) used in our experiments are: 

\begin{itemize}
    \item \textbf{BIOFuzz} \cite{thome2014search}: Search-based tool that generates test cases using context-free grammar based fitness functions. 
    \item \textbf{SQLMap} \cite{sqlmap}: Uses predefined syntax to generate test cases without any active learning component.
    \item \textbf{Baseline Grammar Mutator}: Grammar-based generator with grammar-based mutator created by one of the authors of this paper. 
\end{itemize}

We also compare against a popular white box fuzzer, \textbf{Ardilla} \cite{kieyzun2009automatic}, which uses symbolic execution to find SQLi and XSS vulnerabilities in PHP/MySQL applications. Lastly, we also created two baseline fuzzers: a \textbf{baseline random mutator} that uses a human-written grammar-based generator with random mutations and a \textbf{baseline random fuzzer} that generates random strings inputs. 

\subsection{Benchmarks and Computational Environment}
For a fair comparison with DeepSQLi and SQLmap, we use the same 4 benchmarks as their author \cite{liu2020deepsqli}. Similarly, to compare with BIOFuzz and Ardilla, we use the same 4 benchmarks as their authors \cite{kieyzun2009automatic, thome2014search}. Please refer to their papers for more information about these benchmarks. Furthermore, we created a custom Web application written in PHP with MySQL as a backend database system along with the Flask micro Web framework with SQLite database engine to create small Web applications with SQLi and XSS vulnerabilities. This benchmark comprises different bugs and includes regex sanitizers, which are missing in most of the other benchmarks. We performed our training using an Intel i7 8\textsuperscript{th} Gen 3.20 GHz CPU, 32 GB memory, and 64-bit Ubuntu 18 Desktop. In order to eliminate bias, all the ML-based tools were provided with the same seed input and the same training time on the same hardware. 

\subsection{Metrics}
The metrics used by us have been widely used in this field \cite{bozic2015evaluation, thome2014search, kieyzun2009automatic, liu2020deepsqli} to serve as an important indicator for finding Web application vulnerabilities and establishing the efficacy of fuzzers. We evaluated the above-mentioned tools on the following criteria: 
\begin{itemize}
  \item \textbf{Time to first attack:} Time in seconds to output the first attack vector that exposes a vulnerability.
  \item \textbf{Unique fields:} Number of unique website fields where the tool exposed vulnerabilities.
  \item \textbf{Vulnerabilities found:} Different categories of errors exposed, e.g., unique pairs of input parameters and query statements for SQLi attacks. Attack vectors with INSERT, UNION, or UPDATE are counted as separate categories.
  \item \textbf{Parser Penalties:} Ratio of the number of candidate strings rejected by the parser to the total number of candidate strings generated. 
  \item \textbf{Attack Rate or Error Rate:} Ratio of the number of candidate strings that led to an attack to the total number of candidate strings generated. 
  \item \textbf{Time:} CPU wall clock time in secs to find all vulnerabilities. 
\end{itemize}

\begin{table*}[ht]
\centering
\caption{Comparison results of different ML and non-ML fuzzers}
\label{tab:rq1}
\resizebox{\textwidth}{!}{%
\begin{tabular}{llcccc}
\hline
\multicolumn{1}{c}{\textbf{Models}} & \multicolumn{1}{c}{\textbf{Category}} & \textbf{\begin{tabular}[c]{@{}c@{}}Time to\\ first attack (s)\end{tabular}} & \textbf{\begin{tabular}[c]{@{}c@{}}Unique\\ fields\end{tabular}} & \textbf{\begin{tabular}[c]{@{}c@{}}Vulnerabilities\\ found\end{tabular}} & \textbf{\begin{tabular}[c]{@{}c@{}}Parser\\ Penalties (\%)\end{tabular}} \\ \hline
BERT RL Fuzzer (ours) & Multi-head self-attention PPO & \textbf{102} & \textbf{52} & \textbf{61} & 5\% \\ \hline
DeepFix & \begin{tabular}[c]{@{}l@{}}GRU-based (modified to use with \\ DQN for both SQLi and XSS)\end{tabular} & 224 & 39 & 44 & 39\% \\ \hline
DeepXSS & \begin{tabular}[c]{@{}l@{}}LSTM with word2vec (modified to use \\ with DQN for both SQLi and XSS)\end{tabular} & 267 & 35 & 40 & 45\% \\ \hline
DeepFuzz & \begin{tabular}[c]{@{}l@{}}RNN fuzzer (modified to use \\ with DQN for both SQLi and XSS)\end{tabular} & 254 & 28 & 29 & 41\% \\ \hline
DQN fuzzer & \begin{tabular}[c]{@{}l@{}}Deep Q Network-based \\ fuzzing using DNN for SQLi\end{tabular} & 110 & 9 & 10 & \textbf{0\%} \\ \hline
\begin{tabular}[c]{@{}l@{}}Baseline Grammar Mutator \end{tabular} & Grammar-based generator and mutator & NA & 0 & 0 & \textbf{0\%} \\ \hline
\begin{tabular}[c]{@{}l@{}}Baseline Random Mutator\end{tabular} & Grammar-based generator with random mutation & NA & 0 & 0 & 100\% \\ \hline
Baseline Random Fuzzer & Random & NA & 0 & 0 & 100\% \\ \hline
\end{tabular}%
}
\end{table*}

\section{Evaluation} \label{eval}

Our empirical evaluation of \toolname aims to answer the following research questions:

\textbf{RQ1 (Efficacy and Efficiency of our tool \toolname against State-of-the-art Fuzzers):} How well does \toolname perform against other non-ML and ML-based fuzzers in terms of time to first attack, unique fields, vulnerabilities found, parser penalties, etc.?

\textbf{RQ2 (Ablation Studies):} How does removing/inserting different components of the \toolname impact the tool's performance? 

\textbf{RQ3 (Extensibility of \toolname to Different Categories of Attacks and Victim Applications):} How well does \toolname extend to other kinds of attacks and victim applications?

\subsection{RQ1 (Efficacy/Efficiency against State-of-the-art Fuzzers)}
We compared our tool against different ML and non-ML-based fuzzers, running them on the 9 Web applications with a total timeout of 30 minutes per tool. To avoid bias, all the ML-based tools were provided with the same seed input and the same training time on the same hardware. To make a fair comparison with all the available tools, we had to modify some of these tools (Section \ref{expt_setup}). We evaluated the tools on time to first attack, unique fields, number of vulnerabilities found, and parser penalties. 

We find that \toolname outperforms all the other tools in terms of time to first attack, unique fields, and the number of vulnerabilities found (Table \ref{tab:rq1}). More specifically, on the time-to-first-attack metric, our tool is 54\% faster than the nearest competing tool, finding 17 new vulnerabilities in 13 new unique fields. The only metric where \toolname doesn't outperform all other tools is parser penalties, where it has a score of 5\% and the grammar-adhering mutation fuzzer that we wrote has a parser penalty score of 0\%. This is to be expected since in the case of the grammar-adhering mutation fuzzer, we explicitly specified the grammar, while \toolname learned a representation of the grammar of the victim application with high accuracy.

Further, while the grammar-based mutation fuzzer has a 0\% parsing error, it is not able to find any vulnerabilities in the Web application. The reason is that the space of inputs that the fuzzer has to search over is astronomical, and an unguided search in such contexts is bound to fail in practice. By contrast, observe that \toolname is able to heuristically generate attack vectors, based on the generalization of previously seen patterns, thus reducing the search space dramatically. Note that Web developers are often particularly interested in such attack vectors wherein they may have specified some simple patterns, but missed combinations of such patterns that are also likely to be attack vectors.

DQN fuzzer \cite{zhou2021autonomous, erdodi2021simulating} only predicts the escape characters, adding or deleting column names on two specific grammar-adhering attack patterns for generating SQLi attack vectors. So even though it does preserve the grammar, it can only generate naive union-based and tautology attack vectors. 

We observe that non-ML grammar-based tools are ineffective because they cannot `learn' to create attack vector-provoking mutations, and resort to unguided search. DQN fuzzer, an RL-based fuzzer, is not readily adaptable as one would need to define a typical attack-provoking pattern (e.g., a generic union-based attack string for SQLi) and attack-specific mutation operations corresponding to the specified pattern. Moreover, in the absence of such vulnerabilities, the tool would not be able to identify any new attack vectors. The random fuzzers lead to the highest parser penalties exposing no vulnerabilities because they do not generate grammar-adhering strings. Other ML fuzzers like DeepFuzz, DeepXSS, and DeepFix require a manual effort to create a labeled training dataset making it challenging to adapt to new attacks without prior domain knowledge. Even after using an RL technique (DQN) to bypass the training dataset creation step for the unavailable attacks, most of the candidate strings after mutations are not grammar-aware (as shown by an increased parser penalty). This results in wasting a lot of time getting stuck at the parser. On the other hand, our tool is \textbf{automatic} (learns the mutation operator without human assistance), \textbf{grammar-adhering} (output attack vectors that adhere to the grammar of victim applications with high accuracy), \textbf{effective} (is able to find more security vulnerabilities than competing state-of-the-art tools) and \textbf{efficient} (time to first attack is low). 

\subsection{RQ2 (Ablation Studies)}

\begin{table*}[t]
\centering
\caption{Impact of design choices: Ablation studies}
\label{tab:rq2}
\begin{tabular}{lcccc}
\hline
\multicolumn{1}{c}{\textbf{Models}} & \multicolumn{1}{c}{\textbf{\begin{tabular}[c]{@{}c@{}}Time to\\ first attack (s)\end{tabular}}} & \multicolumn{1}{c}{\textbf{\begin{tabular}[c]{@{}c@{}}Unique\\ fields\end{tabular}}} & \multicolumn{1}{c}{\textbf{\begin{tabular}[c]{@{}c@{}}Vulnerabilities\\ found\end{tabular}}} & \textbf{\begin{tabular}[c]{@{}c@{}}Parser\\ Penalties (\%)\end{tabular}} \\ \hline
\multicolumn{1}{l}{GRU based DQN} & \multicolumn{1}{c}{224} & \multicolumn{1}{c}{39} & \multicolumn{1}{c}{44} & 39\% \\ \hline
\multicolumn{1}{l}{GRU based PPO} & \multicolumn{1}{c}{197} & \multicolumn{1}{c}{39} & \multicolumn{1}{c}{45} & 43\% \\ \hline
\multicolumn{1}{l}{Multi-head self-attention DQN} & \multicolumn{1}{c}{155} & \multicolumn{1}{c}{43} & \multicolumn{1}{c}{48} & 31\% \\ \hline
\multicolumn{1}{l}{Multi-head self-attention PPO} & \multicolumn{1}{c}{141} & \multicolumn{1}{c}{43} & \multicolumn{1}{c}{50} & 28\% \\ 
\multicolumn{1}{l}{+ improved penalties and rewards} & \multicolumn{1}{c}{125} & \multicolumn{1}{c}{48} & \multicolumn{1}{c}{57} & 18\% \\ 
\multicolumn{1}{l}{+ BERT pre-training} & \multicolumn{1}{c}{118} & \multicolumn{1}{c}{\textbf{52}} & \multicolumn{1}{c}{\textbf{61}} & \textbf{5\%} \\ 
\multicolumn{1}{l}{+ MAB seed filtering} & \multicolumn{1}{c}{\textbf{102}} & \multicolumn{1}{c}{\textbf{52}} & \multicolumn{1}{c}{\textbf{61}} & \textbf{5\%} \\ \hline
\end{tabular}
\end{table*}

We use the same Web applications used above and perform an ablation study to observe how every component plays an essential role in \toolname. We observe that using an attention model over a recurrent-based GRU model leads to a quicker first attack, lower parser penalties, and increased vulnerability detection (Table \ref{tab:rq2}). Introducing a PPO agent with improved reward signals (Section \ref{rew_pen}) in the RL loop of \toolname, significantly increased the number of unique fields (+5) and vulnerabilities found (+9). This result shows that using an improved reward signal helps \toolname to better explore the search space. There are also clear improvements over a Deep Q Network (DQN) counterpart because PPO can tackle large action spaces and sparse rewards~\cite{schulman2017proximal}. 

Moreover, using a BERT model as an RL agent led to a drastic decrease in parser penalties (-13\%) relative to a vanilla RL agent. This result shows that using a BERT model enables \toolname to be significantly more grammar-adhering than comparable techniques.

Also, seed input filtering using an MAB agent decreased our time to first attack (-13.56\%). As also observed by BanditFuzz~\cite{scott2021banditfuzz}, single agent RL models can get stuck in local minima and take longer to find an output attack vector~\cite{saavedra2019review, duchene2013fuzz, gerlich2020optimizing, manes2020ankou}. Therefore, using a lightweight secondary agent helps to learn which seed is most likely to lead a successful attack performing better than a deterministic or random choice of selecting an input seed. 


\subsection{RQ3 (Extensibility to Different Categories of Attacks)}
We compare our experiments on two sets of 8 real-world benchmarks (4 per set) against state-of-the-art black-box and white-box tools. To show that our tool can be easily extended to other categories of attacks, other than SQLi, we evaluate on one of the real-world benchmark sets with XSS vulnerabilities present. The ease of pre-training a grammar-adhering model helps our tool extend to the new attack easily.

\begin{table*}[ht]
\centering
\caption{Comparison with respect to DeepSQLi and SQLmap on real-world benchmarks}
\label{tab:rq3_1}
\begin{tabular}{lllllll}
\hline
\multicolumn{1}{c}{\multirow{2}{*}{\textbf{Websites}}} & \multicolumn{2}{c}{\textbf{BERT   RL   Fuzzer}} & \multicolumn{2}{c}{\textbf{DeepSQLi}} & \multicolumn{2}{c}{\textbf{SQLmap}} \\ \cline{2-7} 
\multicolumn{1}{c}{} & \multicolumn{1}{c}{\textbf{Attack   rate}} & \multicolumn{1}{c}{\textbf{Time   (s)}} & \multicolumn{1}{c}{\textbf{Attack   rate}} & \multicolumn{1}{c}{\textbf{Time   (s)}} & \multicolumn{1}{c}{\textbf{Attack   rate}} & \multicolumn{1}{c}{\textbf{Time   (s)}} \\ \hline
Employee & \multicolumn{1}{l}{\textbf{10.64\%}} & \textbf{129} & \multicolumn{1}{l}{8.50\%} & 355 & \multicolumn{1}{l}{4.40\%} & 1177 \\ \hline
Classifieds & \multicolumn{1}{l}{\textbf{11.92\%}} & \textbf{125} & \multicolumn{1}{l}{7.54\%} & 236 & \multicolumn{1}{l}{4.03\%} & 931 \\ \hline
Portal & \multicolumn{1}{l}{\textbf{11.51\%}} & \textbf{130} & \multicolumn{1}{l}{8.55\%} & 357 & \multicolumn{1}{l}{3.56\%} & 2105 \\ \hline
Events & \multicolumn{1}{l}{\textbf{11.05\%}} & \textbf{121} & \multicolumn{1}{l}{9.14\%} & 259 & \multicolumn{1}{l}{4.48\%} & 1094 \\ \hline
\end{tabular}
\end{table*}

\begin{table*}[ht]
\centering
\caption{Comparison with respect to BIOFuzz and Ardilla on real-world benchmarks}
\label{tab:rq3_2}
\begin{tabular}{llllllll}
\hline
\multicolumn{1}{c}{\multirow{2}{*}{\textbf{Websites}}} & \multicolumn{1}{c}{\multirow{2}{*}{\textbf{Mode}}} & \multicolumn{2}{c}{\textbf{BERT RL Fuzzer}} & \multicolumn{2}{c}{\textbf{BIOFuzz}} & \multicolumn{2}{c}{\textbf{Ardilla}} \\ \cline{3-8} 
\multicolumn{1}{c}{} & \multicolumn{1}{c}{} & \multicolumn{1}{c}{\textbf{\# Vul}} & \multicolumn{1}{c}{\textbf{\begin{tabular}[c]{@{}c@{}}Time\\ (s)\end{tabular}}} & \multicolumn{1}{c}{\textbf{\# Vul}} & \multicolumn{1}{c}{\textbf{\begin{tabular}[c]{@{}c@{}}Time\\ (s)\end{tabular}}} & \multicolumn{1}{c}{\textbf{\# Vul}} & \multicolumn{1}{c}{\textbf{\begin{tabular}[c]{@{}c@{}}Time\\ (s)\end{tabular}}} \\ \hline
\multirow{3}{*}{Webchess} & SQLi & \multicolumn{1}{l}{\textbf{13}} & \textbf{300} & \multicolumn{1}{l}{\textbf{13}} & 596 & \multicolumn{1}{l}{12} & 1800 \\ \cline{2-8} 
 & XSS1 & \multicolumn{1}{l}{\textbf{13}} & \textbf{300} & \multicolumn{1}{l}{-} & - & \multicolumn{1}{l}{\textbf{13}} & 1800 \\ \cline{2-8} 
 & XSS2 & \multicolumn{1}{l}{0} & 300 & \multicolumn{1}{l}{-} & - & \multicolumn{1}{l}{0} & 1800 \\ \hline
\multirow{3}{*}{Schoolmate} & SQLi & \multicolumn{1}{l}{\textbf{6}} & \textbf{1000} & \multicolumn{1}{l}{\textbf{6}} & 1687 & \multicolumn{1}{l}{\textbf{6}} & 1800 \\ \cline{2-8} 
 & XSS1 & \multicolumn{1}{l}{\textbf{10}} & \textbf{1000} & \multicolumn{1}{l}{-} & - & \multicolumn{1}{l}{\textbf{10}} & 1800 \\ \cline{2-8} 
 & XSS2 & \multicolumn{1}{l}{1} & 1000 & \multicolumn{1}{l}{-} & - & \multicolumn{1}{l}{\textbf{2}} & 1800 \\ \hline
\multirow{3}{*}{FAQForge} & SQLi & \multicolumn{1}{l}{\textbf{1}} & 44 & \multicolumn{1}{l}{\textbf{1}} & \textbf{32} & \multicolumn{1}{l}{\textbf{1}} & 1800 \\ \cline{2-8} 
 & XSS1 & \multicolumn{1}{l}{\textbf{4}} & \textbf{120} & \multicolumn{1}{l}{-} & - & \multicolumn{1}{l}{\textbf{4}} & 1800 \\ \cline{2-8} 
 & XSS2 & \multicolumn{1}{l}{0} & 120 & \multicolumn{1}{l}{-} & - & \multicolumn{1}{l}{0} & 1800 \\ \hline
\multirow{3}{*}{geccbblite} & SQLi & \multicolumn{1}{l}{\textbf{4}} & \textbf{300} & \multicolumn{1}{l}{\textbf{4}} & 656 & \multicolumn{1}{l}{2} & 1800 \\ \cline{2-8} 
 & XSS1 & \multicolumn{1}{l}{0} & 300 & \multicolumn{1}{l}{-} & - & \multicolumn{1}{l}{0} & 1800 \\ \cline{2-8} 
 & XSS2 & \multicolumn{1}{l}{1} & 300 & \multicolumn{1}{l}{-} & - & \multicolumn{1}{l}{\textbf{4}} & 1800 \\ \hline
\end{tabular}
\end{table*}

The first set of benchmarks consists of six real-world commercial Web applications popularly used by researchers \cite{halfond2006using, liu2020deepsqli}. These Web applications are written in Java and use a MySQL back-end database. We compare our tool \toolname against two black-box approaches DeepSQLi~\cite{liu2020deepsqli} and SQLMap~\cite{sqlmap}. We reuse the results reported by DeepSQLi \cite{liu2020deepsqli} and perform our experiments, repeating it 20 times using the same compute setup used by DeepSQLi \cite{liu2020deepsqli}. We evaluate our tool on the same metrics they used, i.e., Attack Rate (or Error Rate) and CPU wall clock time. We found that our tool significantly outperforms both the tools achieving a higher error rate, 1.91-4.38\% more than the nearest competing tool, in less than half the time taken while discovering all the vulnerabilities reported by the tool authors (Table \ref{tab:rq3_1}). 

For the second set of benchmarks (Table \ref{tab:rq3_2}), we compared our tool against a popular white-box tool Ardilla \cite{kieyzun2009automatic} and a black-box evolutionary testing-based tool BIOFuzz \cite{thome2014search}. We use the same case studies (benchmark set) as used by Ardilla and BIOFuzz, reusing the results reported by the authors, as Ardilla is not publicly available and BIOFuzz is severely out-of-date. We also evaluated using the same metrics used by the authors of Ardilla, i.e., the number of vulnerabilities detected and run time or timeout in seconds. Ardilla is a relatively old tool and was run on a 30 minutes timeout by the authors for all the case studies. We used the same experimental setup as BIOFuzz for a fair comparison with the tool.  

Our tool can find all the SQLi and XSS1 (first-order XSS) vulnerabilities reported by the authors. Since BIOFuzz only supports SQLi, the authors did not report any XSS1 or XSS2 (second-order XSS) vulnerabilities. Our tool also found three new SQLi vulnerabilities not reported by Ardilla but reported by BIOFuzz. This shows that our tool can find different attack patterns for SQLi vulnerabilities. On the other hand, our tool could not find four XSS2 vulnerabilities reported by Ardilla. Second-order XSS attacks (XSS2) are challenging to find because a sequence of inputs is responsible for creating such an attack. So, an SMT solver-based white box fuzzers like Ardilla can easily infer these test strings. Moreover, our tool is significantly faster (40-60\% improvement) in all the cases except one where the BIOFuzz tool detects an SQLi attack a few seconds faster. Therefore, we can say that our tool can be easily extended to new victim applications as well as a different class of vulnerabilities (e.g., XSS) and can be as effective as a state-of-the-art white-box fuzzer such as Ardilla in finding vulnerabilities in significantly less time. 

\section{Threats to Validity} \label{threats}

\noindent {\bf Validity of Experimental Evaluation:} We compare our tool against 13 other state-of-the-art tools on 9 large real-world benchmarks often used by the authors of competing tools~\cite{liu2020deepsqli, kieyzun2009automatic, thome2014search}. We also use the metrics that have been widely used in this field~\cite{bozic2015evaluation, thome2014search, kieyzun2009automatic, liu2020deepsqli} to serve as an important indicator for finding Web application vulnerabilities. To the best of our knowledge, our experimental evaluation is the most comprehensive and thorough of any fuzzing tool of its kind.

\vspace{0.2cm}
\noindent {\bf Learning General Semantic Structure of Web application vulnerabilities:} As mentioned earlier, BERT models learn some representation of the grammar of their inputs with high accuracy. This of course means that we do not expect our model to learn the grammar perfectly. Having said that, language models \cite{bommasani2021opportunities} such as BERT have had tremendous empirical success when learning sophisticated grammar by leveraging the attention mechanism \cite{vaswani2017attention}. Some examples include code translations from one programming language to another \cite{lachaux2020unsupervised, mastropaolo2021studying}, program synthesis \cite{allamanis2013mining,chen2021evaluating}, and code understanding \cite{mou2016convolutional,guo2020graphcodebert,feng2020codebert}. In all of these applications, an ML model with a highly accurate empirical model of the grammar is a prerequisite. \toolname presents yet another application of BERT that demonstrates its ability to learn a highly accurate model of complex grammar. In evaluation, we observed \toolname could be up to 95\% accurate and is surpassed only by hand-written grammar-adhering mutation fuzzers.

\vspace{0.2cm}
\noindent {\bf Extensibility of \toolname:} In this paper, we extensively evaluated our tool on two orthogonal use cases (i.e., SQLi and XSS). As mentioned above, the ability of language models, such as BERT, to learn empirically accurate representations of sophisticated non-trivial grammars (e.g., programming languages) suggests that our \toolname can be easily extended to other classes of applications and attack vectors. The fuzzing process of \toolname is easily modifiable, customizable, and maneuverable. The designer can modify the tool by providing a supervised set of training examples with specific mutation operations commonly known for the attack patterns the designer wants to focus on. These developer-defined mutation patterns serve as an initial fine-tuning step before starting the RL loop, helping the model to learn better attack patterns faster. More precisely, the fuzzer can be extended to a different application by simply replacing the seed inputs, the pre-trained BERT model, and the environment (program-under-test).

\section{Related Work}
In the domain of fuzzing, Reinforcement Learning has been a popular choice in creating a variety of fuzzing algorithms, especially for mutation operation selection~\cite{bottinger2018deep}. Moreover, penetration testing problems have been previously modeled as RL problems with various abstractions of the problem \cite{sarraute2013penetration, ghanem2020reinforcement, zennaro2020modeling}. SARSA (State-action-reward-state-action) was a popular choice in the early days for this task \cite{becker2010autonomic}, followed by Q-Learning \cite{fang2018emulation}. After the popularity of Deep Learning, Deep Q-Learning was used by Bottinger et al. \cite{bottinger2018deep} for mutation operation selection and by Kuznetsov et al. \cite{kuznetsov2019automated} for exploitability analysis. Drozd et al. \cite{drozd2018fuzzergym} used another variant of Deep Q-Learning called Deep Double Q-Learning along with Long Short-Term Memory (LSTM), a popular model in the field of Natural Language Processing, for mutation operation selection. Similarly, $\mu$4SQLi \cite{appelt2014automated} also performs mutation operations on seed test inputs but is designed for a fixed set of patterns. To tackle the exploitation problem that can often plague RL-based methods, some recent work has focused on exploitation in a simplified SQL environment for specific attack patterns and mutation operators~\cite{erdodi2021simulating, verme2021sql}. 

General purpose fuzzers like AFL \cite{afl}, and PerfFuzz \cite{lemieux2018perffuzz} that are built around bit-string manipulation are not grammar-adhering and hence are unable to produce well-formed inputs for complex grammars typical for web applications. 

BanditFuzz~\cite{scott2021banditfuzz} is an RL-guided performance fuzzer for Satisfiability Modulo Theories (SMT) solvers. However, the RL agent (MAB) is stateless \cite{vermorel2005multi}. The rewards are learned only based on actions, irrespective of the current state (i.e., the seed string, in our case). Further, unlike \toolname, BanditFuzz cannot learn any representation of the grammar of victim applications, and therefore is not grammar-adhering or easily extensible. 

Unlike the above-mentioned approaches to ML/RL and non-ML-based fuzzing, we use a combination of BERT model as the RL agent in \toolname. This enables our tool to be grammar-adhering, extensible, and automatic in a way that is not the case with state-of-the-art fuzzers. Further, our extensive experiments over a large real-world benchmark suite demonstrate that our tool is more effective and efficient than competing tools.

\section{Conclusion}

In this paper, we present \toolname, a Reinforcement Learning (RL)-based fuzzer. \toolname is the first Machine Learning-based fuzzer that uses a BERT architecture and RL-based algorithm without needing a manually crafted grammar file or labeled training dataset. Via an extensive, comprehensive, and thorough empirical evaluation against 13 fuzzers on 9 different benchmarks, we show that our tool \toolname is automatic, extensible, grammar-adhering, efficient, and effective. Our tool is most useful to application developers who find that writing grammar-adhering mutation fuzzers by hand is time-consuming, error-prone, and expensive. Further, our tool is particularly effective for scenarios where application developers may have written sanitizers based on simple attack vectors, missing complex combinations. By contrast, given a rich enough data set of simple attack vectors, \toolname is able to learn complex combinations of attack patterns, thus finding weaknesses in sanitizers in an efficient heuristic way. To the best of our knowledge, no other ML-based fuzzer uses a BERT architecture and an RL-based algorithm to solve this problem of requiring fuzzer users to somehow modify fuzzers to be grammar-adherent to victim applications or requiring them to provide attack-vector patterns, a complex, expensive, and error-prone process. Further, we do not know of any other fuzzer that is extensible without requiring human intervention. In future work, Monte Carlo tree search (MCTS) \cite{browne2012survey} can be used to explore the search space better, similar to how they are used by popular self-play engines \cite{silver2018general}. Additionally, our approach can be applied to assess various types of software that rely on structured inputs, including but not limited to compilers, SMT solvers, and PDF readers.

\bibliography{aaai24}

\end{document}